# MIC: Medical Image Classification Using Chest X-ray (COVID-19 and Pneumonia) Dataset with the Help of CNN and Customized CNN

Nafiz Fahad[†]
Faculty of Information Science and Technology
Multimedia University
Melaka, Malaysia
fahadnafiz1@gmail.com

Fariha Jahan
Department of Computer Science and Engineering
American International University-Bangladesh
Dhaka, Bangladesh
fariha.rainy23@gmail.com

Md Kishor Morol
Department of Computing Science and Information Science
Cornell University
New York, USA
kishoremorol@gmail.com

Rasel Ahmed
Department of Computer Science and Engineering
American International University-Bangladesh
Dhaka, Bangladesh
raselahmed1337@gmail.com

Md. Abdullah-Al-Jubair
Department of Computer Science
American International University-Bangladesh
Dhaka, Bangladesh
abdullah@aiub.edu

## ABSTRACT

The COVID-19 pandemic has had a detrimental impact on the health and welfare of the world's population. An important strategy in the fight against COVID-19 is the effective screening of infected patients, with one of the primary screening methods involving radiological imaging with the use of chest X-rays. Which is why this study introduces a customized convolutional neural network (CCNN) for medical image classification. This study used a dataset of 6432 images named Chest X-ray (COVID-19 & Pneumonia), and images were preprocessed using techniques, including resizing, normalizing, and augmentation, to improve model training and performance. The proposed CCNN was compared with a convolutional neural network (CNN) and other models that used the same dataset. This research found that the Convolutional Neural Network (CCNN) achieved 95.62% validation accuracy and 0.1270 validation loss. This outperformed earlier models and studies using the same dataset. This result indicates that our models learn effectively from training data and adapt efficiently to new, unseen data. In essence, the current CCNN model achieves better medical image classification performance, which is why this CCNN model efficiently classifies medical images. Future research may extend the model's application to other medical imaging datasets and develop real-time offline medical image classification websites or apps.

## 1 INTRODUCTION

A crucial stage in the analysis of medical images is medical image classification, which makes use of many elements to distinguish between distinct medical images, such as imaging modalities or clinical data. Clinicians may be able to assess medical images more swiftly and accurately with the aid of reliable medical image classification. Recent advances in convolutional neural networks (CNNs) have yielded significant benefits for the healthcare industry. The application of artificial intelligence-based computer-aided diagnostic tools in medical environments has been the subject of extensive research due to these improvements. To achieve effective diagnostic performance in medical areas, CNNs can train robust discriminative representations from large volumes of medical data. These methods achieve results similar to those of clinicians and validate their adequate prediction abilities [1-3].

Furthermore, the advent of deep learning in computer vision has made it possible to perform image classification tasks with previously unheard-of levels of accuracy. One of the biggest issues with medical imaging that focuses on inspection, diagnosis, and therapy is classification. Advancements in computing hardware and methodologies have made it possible for computer-aided solutions to possibly assess the clinical judgment of doctors. Several advanced deep convolutional neural network (CNN) models have been created and assessed to categorize radiological images according to illness or histology images. This involves recognizing regions containing normal and malignant cells [4-5].

However, many researchers have classified medical images using deep learning. One approach, which addresses issues with feature extraction and uncertainty quantification, uses a dynamic multiscale convolutional neural network (DM-CNN) to improve medical image classification. The dynamic multiscale feature fusion module (DMFF), hierarchical dynamic uncertainty quantification attention (HDUQ-Attention), multiscale fusion pooling method (MF pooling), and multiobjective loss (MO loss) are the four main components of the DM-CNN, which is a novel convolutional neural network architecture. Together, these parts choose the right convolution kernels for different feature map levels, modify attention weights, utilize Monte Carlo dropout to quantify uncertainty, and balance the trade-offs between classification accuracy and speed. During testing, the DM-CNN achieved state-of-the-art classification accuracy on four medical datasets: dermatology, histopathology, respiratory, and ophthalmology. It performed better than the other models. The model's ability to quantify uncertainty effectively and sustain stable performance across many medical domains is critical for clinical applications, as demonstrated by the results. Despite using sophisticated pooling techniques, the study admits that it is limited in its ability to balance feature representation across various scales and may overfit. This implies that additional refining is necessary to maximize these features while maintaining the high accuracy and efficiency of the model [6].

Another approach is to evaluate various fine-tuning strategies for applying pretrained convolutional neural networks (CNNs) to different medical imaging tasks. They assessed eight fine-tuning approaches—full fine-tuning, linear probing, and gradual unfreezing—across three CNN architectures (ResNet-50, DenseNet-121, and VGG-19) and multiple medical domains (e.g., X-ray, MRI, and histology). The methods were tested on multiple datasets, resulting in varying performance enhancements: some strategies improved accuracy by up to 11% in certain modalities. Compared with other architectures, DenseNet often benefits more distinctively from nontraditional fine-tuning. The results demonstrated the effectiveness of adaptive fine-tuning methods, particularly in contexts where direct transfer learning falls short due to differences between the source and medical image characteristics. However, the study's limitations include potential biases in dataset distribution and the computational cost associated with fine-tuning techniques, which might limit practical application in real-world scenarios where computational resources are constrained [7].

An alternative approach is HiFuse, a unique hierarchical multiscale feature fusion network designed for medical image classification. Adaptive hierarchical feature fusion blocks inverted residual multilayer perceptrons, global and local feature blocks, and other components are used by HiFuse to include global and local features at different levels and improve classification accuracy. With classification accuracies of 85.85% and 86.12% on the ISIC2018 and Kvasir datasets, respectively, this strategy helps to improve the semantic richness of the extracted features and enhances accuracy across multiple medical datasets, including dermatological and histology datasets. The study notes that, in spite of its great accuracy, it is difficult to maintain balance in feature representation across different scales, which may result in overfitting of the model. Nonetheless, the model's performance holds true across several datasets, demonstrating its efficacy in the classification of medical images [8].

As a result, only a small number of studies—like Ting et al. (2022)—have used the same dataset as this one. Their goal was to improve COVID-19-XR classification accuracy by implementing CNN architectures with more layers. Still, the model's accuracy was only 94% [9]. With a validation accuracy of 93.97% for multiclass instances, Patil and Narawade (2024) employed DL models to help with the diagnosis, treatment, and early identification of respiratory disorders such pneumonia and COVID-19 [10]. Elkamouny and Ghantous (2022) utilized eight deep learning models in addition to pretrained models for diagnosis; of these, the Inception-ResNet-v2 model outperformed the others with an accuracy of 95.3% [11]. In Patel S.'s evaluation of DenseNet (2021) for COVID-19 chest X-ray image classification, DenseNet201 was used, and the model's validation accuracy was 93.67% [12]. Even with this earlier research, there are still certain restrictions. Using the same datasets as earlier research, this paper presents a customized convolutional neural network (CCNN) and evaluates it against a standard CNN. The suggested CCNN performs better than rival models, according to the results.

The rest of the article is organized as follows: The study's methodology is thoroughly explained in Section 2, its findings are discussed in Section 3, and its conclusions are discussed in Section 4.

## 2 METHODS

For this research purpose, CNNs and customized CNNs are appropriate.

### 2.1 Proposed Method

The diagram in Figure 1 shows the proposed methods of this study.

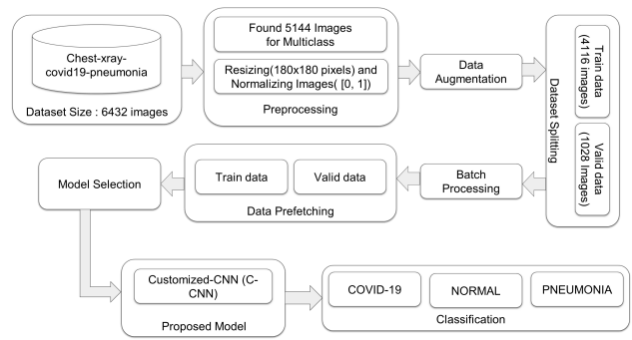

Figure 1: Proposed method of this current study.

## 2.2 Dataset Collection

This study utilized Kaggle's Chest X-ray dataset, specifically focusing on cases with COVID-19 and pneumonia. The dataset comprises chest X-rays of individuals diagnosed with COVID-19, pneumonia, as well as those who are considered normal. The dataset is divided into two folders: "train" and "test". Every folder has three subordinate folders: "COVID-19", "PNEUMONIA", and "NORMAL". The collection comprises 6432 X-ray pictures, of which 20% are designated as test data [13]. Additionally, an image from the dataset is attached below, which is not preprocessed.

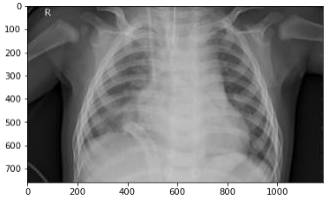

Figure 2: A pneumonia sample before preprocessing

## 2.3 Data Preprocessing

For this study, data preprocessing is a critical step in building effective machine learning models, especially when dealing with image data such as medical images. There are several common data preprocessing techniques used in the context of medical image analysis, as illustrated below.

*2.2.1 Resizing and Normalizing Images*

For the present study, images were resized to a consistent shape (180x180 pixels) to ensure uniformity across all inputs. This is crucial for the model to process them efficiently. Normalization is achieved by rescaling the values of each pixel to fit within the range of 0 to 1. This helps the acceleration of convergence during training by assuring that all features (pixel values) participate equally [14].

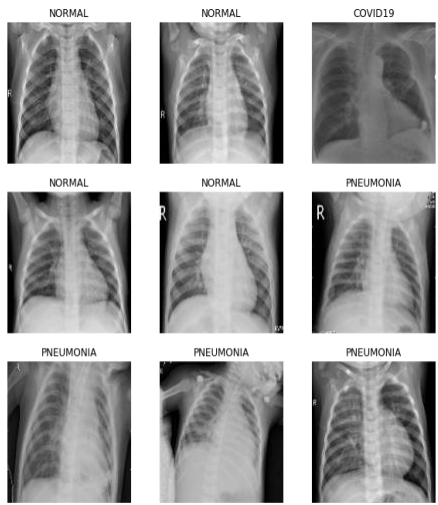

Figure 3: **First 9 images after resizing and normalization of the dataset.**

*2.2.2 Data Augmentation*

The present study applied a data augmentation approach to artificially increase the dataset's size by generating modified copies of the images inside the dataset. Additionally, images are subjected to random horizontal flipping in order to replicate various orientations. Images are randomly rotated by a small angle (5% of 360 degrees). Random Zoom: The images are zoomed in or out by a certain percentage (10%). However, augmentation helps in building robust models by introducing variability in the training data, which can improve generalization [15]. Additionally, Figure 3 shows the first 9 images after data augmentation.

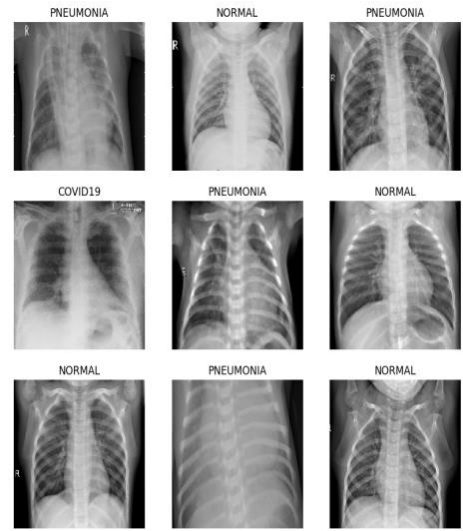

Figure 4: **First 9 images after data augmentation of the dataset.**

*2.2.3 Dataset Splitting*

The dataset is often divided into training and validation sets for this kind of investigation. This allows the model to be trained on one set of data, refined on another, and finally tested on data that hasn't been seen yet. Usually, 20% of the training set is utilized for validation, which allows us to track the model's performance as it is being trained. A 20% training dataset was also employed in this investigation.

*2.2.4 Batch processing*

The data used in the present study were processed in batches (32 images per batch). This is a practical approach for making efficient use of memory and improving computational speed. Batching also allows the model to update its weights after seeing several examples and tends to make the learning process more stable.

*2.2.5 Prefetching*

This study used the prefetching technique to preload data into buffer memory to speed up the training process. This technique ensures that the next batch of data is readily available for the model to process as soon as it finishes with the current batch, reducing the time spent waiting for data loading.

## 2.4 Model Selection

Model selection is an important stage. In the present study, a convolution neural network (CNN) and a customized convolution neural network (CCNN) were used.

*2.4.1 CNN*
*The* CNN model's structure includes several layers that progressively process and simplify the image input. Initially, an image augmentation step enhances the input data to help the model generalize better from the training data. Following this, the network employs multiple layers that apply filters and reduce the size of the data, interspersed with dropout layers that randomly ignore parts of the data during training to prevent the model from memorizing the training data too closely.

The network complexity increases by adding more filters in deeper layers. After processing through these layers, the data are flattened and passed through a dense layer for final classification. The configuration of the output layer depends on whether the model is classified into two categories (using sigmoid activation) or more (using softmax activation).

On the other hand, the model employs learning rate scheduling, which modifies the model's learning process as training advances, and early stopping to prevent overfitting. The model learns to classify new images using the patterns it has acquired during training, which is carried out using predetermined epochs. Additionally, the CNN model architecture's pseudocode is provided below.

---

**Pseudocode of CNN model architecture**
function MAKE_MODEL(input_shape, num_classes):
   inputs ← keras.Input(shape=input_shape)
      x ← data_augmentation(inputs)
      x ← layers.experimental.preprocessing.Rescaling(1.0/255)(x)
      x ← apply several convolutional, activation, and pooling layers
      x ← layers.Flatten()(x)
      x ← layers.Dense(256, activation='relu')(x)
      if num_classes = 2 then
        activation ← "sigmoid"
         units ← 1
      else
         activation ← "softmax"
         units ← num_classes
       x ← layers.Dropout(0.5)(x)
      outputs ← layers.Dense(units, activation=activation)(x)
      return keras.Model(inputs, outputs)
   end function

---

*2.4.2 CCNN*

To classify images, a CCNN model is constructed with the help of the Keras package. The function `make_model` defines the model and requires two parameters, `input_shape} and `num_classes}. Without gathering new data, it begins by taking in input images and utilizing data augmentation to increase the variety of training instances. Next, the resampled images are assigned values ranging from 0 to 1, hence increasing the training efficiency.

The model's core consists of many convolutional layers, each of which is separated from the others by batch normalization, activation of rectified linear units (ReLUs), max pooling to minimize dimensionality, and dropout to reduce overfitting. The architecture is clearly customized, with higher dropout rates and different filter sizes applied to different layers in an effort to maximize performance.

The global average pooling layer, which comes after the convolutional layers in the model, aids in lowering the overall number of parameters and making the model simpler. To further prevent overfitting, a dense layer with ReLU activation is added, and then there is another dropout, but this time at a rate of 0.5.

The model's output layer varies according to the number of classes: a sigmoid activation is used for multiclass classification, while a softmax activation is used for binary classification (num_classes == 2}). Because of its adaptability, the model can change depending on the classification task.

The Adam optimizer, which uses a modest learning rate to guarantee gradual and steady learning, is used to generate the model. It makes use of a sparse categorical cross-entropy loss function. To improve efficiency, the training technique includes callbacks, such as early stopping, to terminate training if the validation loss does not grow. Furthermore, a learning rate scheduler with adjustable decay rates is used to progressively lower the learning rate following a predetermined number of epochs.

Basically, this model is tailored to perform better and more efficiently for image classification tasks by making specific changes to its architecture and training plan. It is evident from the highlighted setup and customization locations where certain changes have been made for possible improvement [16].

Furthermore, the CCNN is intended to improve the categorization of medical images, specifically for the identification of pneumonia and COVID-19 from chest X-ray images. Advanced data augmentation methods that increase the diversity of training instances and boost the generalization of the model are used by the CCNN. These methods include random horizontal flipping, small-angle rotations, and zooming. Rescaling images to values between 0 and 1 increases the effectiveness of the training process. Multiple convolutional layers with batch normalization and ReLU activation make up the core of the CCNN. Filter sizes of 32, 64, 128, and 256 are used to capture varying levels of information. Dropout layers with rates rising from 0.3 to 0.5 stop overfitting, whereas max pooling layers lower dimensionality and computing burden. A global average pooling layer minimizes the overall number of parameters, simplifying the model and lowering the danger of overfitting. The output layer uses sigmoid or softmax activation to adjust to binary or multi-class classification tasks, while the dense layer with 256 units and ReLU activation processes the flattened feature maps. The CCNN performs better as a result of these architectural decisions since they allow it to extract strong features, reduce overfitting, and adapt effectively to various classification tasks. With a validation accuracy of 95.62% and a validation loss of 0.1270, the CCNN has outperformed other models in terms of learning from training data and reliably classifying newly discovered data. Additionally, the CCNN model architecture's pseudocode is provided below.

**Pseudocode of CNN model architecture**
```
function MAKE_MODEL(input_shape, num_classes):
    inputs ← keras.Input(shape=input_shape)
    x ← data_augmentation(inputs)
    x ← layers.experimental.preprocessing.Rescaling(1.0/255)(x)

    # Adding more convolutional layers and adjusting filter sizes
    for filters in [32, 64, 128, 256]:
        x ← layers.Conv2D(filters, 3, padding="same", kernel_regularizer=l2(0.01))(x)
        x ← layers.BatchNormalization()(x)
        x ← layers.Activation("relu")(x)
        x ← layers.MaxPooling2D()(x)
        x ← layers.Dropout(0.3)(x)  # Increased dropout rate

    x ← layers.GlobalAveragePooling2D()(x)   # Using GlobalAveragePooling2D before the dense layer
    x ← layers.Dense(256, activation='relu', kernel_regularizer=l2(0.01))(x)
    x ← layers.Dropout(0.5)(x)

    if num_classes = 2 then
        activation ← "sigmoid"
        units ← 1
    else
        activation ← "softmax"
        units ← num_classes

    outputs ← layers.Dense(units, activation=activation)(x)
    return keras.Model(inputs, outputs)
end function
```

## 2.5 Evaluation Metric

For the current study, models are evaluated by an evaluation metric named accuracy. Accuracy is the most often used metric for precisely determining whether something is true or untrue [17, 18]. The formula is as follows:

$$Accuracy = \frac{TP+FP}{TP+FP+TN+FN} \qquad (10)$$

Here, $TP$ = true positive, $FP$ = false positive, $TN$ = true negative, and $FN$ = false negative.

## 3 RESULTS

### 3.1 CNN training and validation loss

Figure 5 shows the training and validation loss of a convolutional neural network (CNN) over the first 24 epochs, which is part of a training regimen with a total of 50 epochs. Initially, both the training and validation losses are high, starting at approximately 0.89. Rapid improvement is observed by the 5th epoch, with the training loss significantly reduced to approximately 0.27 and the validation loss to approximately 0.17. However, there is a notable spike in the validation loss to 0.51 by the 8th epoch, suggesting potential overfitting at that stage. After this spike, the losses gradually stabilize, with the training loss being consistently lower than the validation loss. By the 20 h epoch, the training loss has further decreased to approximately 0.14, indicating effective learning and fitting to the training data, while the validation loss has also decreased and leveled off, reflecting better generalization to new data.

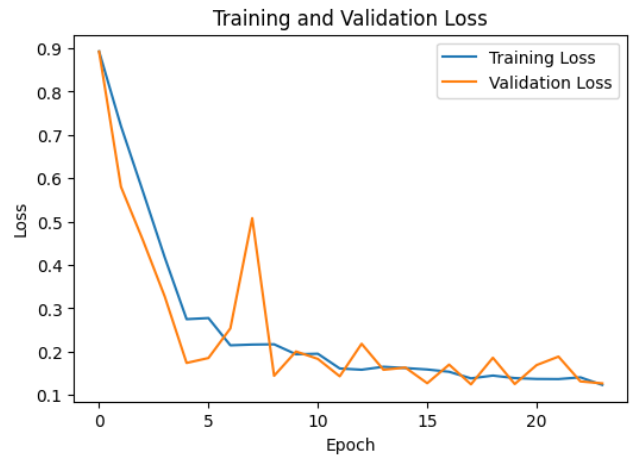

Figure 5: **Training and validation loss of CNN.**

### 3.2 CNN training and validation accuracy

The first 20 epochs of CNN model training and validation accuracy are shown in the figure. The training accuracy starts at 63.22% and improves quickly, reaching above 95.53% by the 24th epoch. The validation accuracy begins at 66.34% and follows a similar upward trend but exhibits more volatility, which is particularly noticeable around the 6th and 8th epochs, where it dips before rising again. This fluctuation might indicate moments when the model struggled with new data before adapting. By the 24th epoch, both accuracies are high, with 95.53% validation accuracy, which is slightly below the training accuracy of 95.85%, suggesting that while the model is performing well, there is still room for improvement.

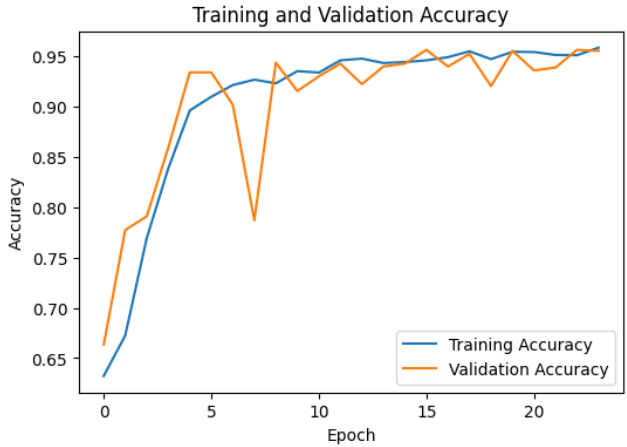

Figure 6: **Training and validation accuracy of CNN.**

### 3.3 CCNN training and validation loss

Figure 7 shows the training and validation loss of the C-CNN model over 24 epochs. Both losses start high (over 11) and decrease steadily, indicating that the model is learning from the data. The losses closely follow each other, which suggests good generalizability without significant overfitting. However, there is a notable increase in the validation loss at the 20th epoch, which might indicate that the model is starting to overfit. By the 24th epoch, both losses decreased significantly, with the training loss being slightly lower than the validation loss. This shows that the model's performance has greatly improved throughout the training process.

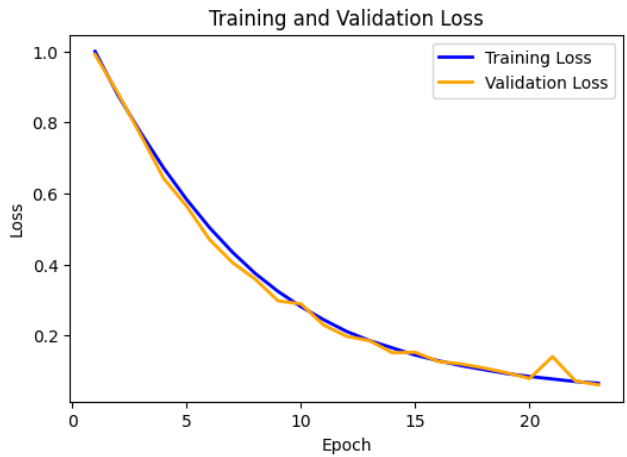

Figure 7: **Training and validation loss of CCNN.**

### 3.4 C-CNN Training and Validation Loss

Figure 8 presents the training and validation accuracy of a convolutional neural network model over 24 epochs. The training accuracy starts from a lower value but rapidly increases, stabilizing at approximately 90% to 95% for the majority of the training. This indicates that the model consistently performs well on the training data throughout the later epochs. In contrast, the validation accuracy decreases but significantly improves as the number of epochs increases, suggesting that the model gradually learns to generalize better to unseen data. However, the validation accuracy shows considerable variability, particularly a sharp decline around the 20th epoch, which suggests potential issues such as overfitting or an anomaly in the validation data at that point. Despite this, both accuracies reach high values by the final epochs (training accuracy 95.70% and validation accuracy 95.62%), demonstrating overall successful learning and adaptation by the model.

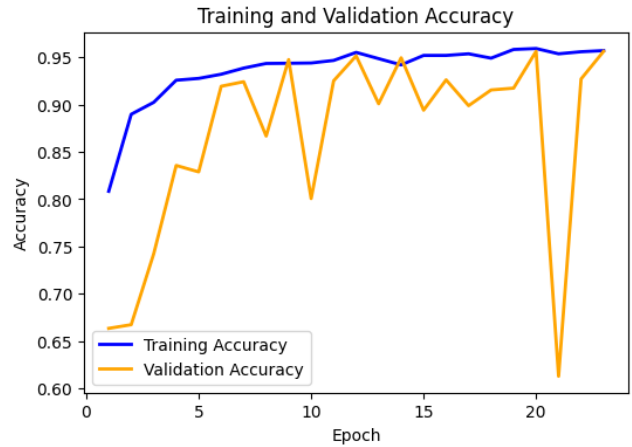

Figure 8: **Training and validation accuracy of CCNN.**

### 3.5 Comparison

Table I presents a comparison between prior studies and the current study. According to the data, the suggested model in this study showed better results compared to the other models in terms of validation accuracy. The validation loss is a metric that quantifies the performance of a model on the validation set. It is calculated by applying the model's loss function to the validation data and averaging the results. Lower validation loss values indicate that the model is predicting the validation data more accurately.

Table I: Comparison between previous and current studies

| Reference | Model | Validation Accuracy | Validation Loss |
| --- | --- | --- | --- |
| [10] Patil and Narawade (2024) | Deep CNN | 0.9397 | 0.3719 |
| [9] Ting et al.(2022) | CNN architectures with additional layers | 0.94 | Not provided |
| [11] Elkamouny and Ghantous (2022) | Inception-ResNet-v2 | 0.953 | Not provided |

| [12] Patel, S. (2021) | DenseNet201 | 0.9367 | 0.1653 |
|---|---|---|---|
| **Proposed** | **Customized-CNN (C-CNN)** | **0.9562** | **0.1270** |

## 4  CONCLUSIONS

This work represents a notable advancement in the field of medical image classification. The performance of the customized convolutional neural network (CCNN) developed herein is validated with an accuracy level of 95.62%. This CCNN model outperforms the CNN model and other existing studies that used the same dataset. This study used rigorous data preprocessing steps (resizing, normalizing, and augmenting), training and tuning strategies to solve the challenges of medical image classification. The customized model could still be optimized by expanding to other larger medical imaging datasets and more diverse datasets. This study contributes valuable insights to the academic field. This study also contributes to significant improvements in healthcare and helps to efficiently classify medical images.

## ACKNOWLEDGMENTS

This research is funded by Elitelab.ai. For the support and fund the authors give a special thanks to Elitelab.ai.

## REFERENCES

[1] Chung-Ming Lo and Peng-Hsiang Hung. 2022. Computer-aided diagnosis of ischemic stroke using multidimensional image features in carotid color Doppler. *Computers in Biology and Medicine* 147: 105779. http://doi.org/10.1016/j.compbiomed.2022.105779

[2] Weiming Hu, Chen Li, Xiaoyan Li, Md Mamunur Rahman, Jiquan Ma, Yong Zhang, Haoyuan Chen, Wanli Liu, Changhao Sun, Yudong Yao, Hongzan Sun, Marcin Grzegorzek. 2022. GasHisSDB: A new gastric histopathology image dataset for computer aided diagnosis of gastric cancer. *Computers in Biology and Medicine* 142: 105207. http://doi.org/10.1016/j.compbiomed.2021.105207

[3] Qiongjie Hu, Chong Chen, Shichao Kang, Ziyan Sun, Yujin Wang, Min Xiang, Hanxiong Guan, Liming Xia, Shaofang Wang. 2022. Application of computer-aided detection (CAD) software to automatically detect nodules under SDCT and LDCT scans with different parameters. Computers in Biology and Medicine 146: 105538. http://doi.org/10.1016/j.compbiomed.2022.105538

[4] Eirini Arvaniti, Kim S. Fricker, Michael Moret, Niels Rupp, Thomas Hermanns, Christian Fankhauser, Norbert Wey, Peter J. Wild, Jan H. Rüschoff, Manfred Claassen. 2018. Automated Gleason grading of prostate cancer tissue microarrays via deep learning. *Scientific Reports* 8, 1. http://doi.org/10.1038/s41598-018-30535-1

[5] Okyaz Eminaga, Mahmoud Abbas, Jeanne Shen, Mark Laurie, James D. Brooks, Joseph C. Liao, Daniel L. Rubin. 2023. PlexusNet: A Neural Network architectural concept for medical image classification. *Computers in Biology and Medicine* 154: 106594. http://doi.org/10.1016/j.compbiomed.2023.106594

[6] Qi Han, Xin Qian, Hongxiang Xu, Kepeng Wu, Lun Meng, Zicheng Qiu, Tengfei Weng, Baoping Zhou, Xianqiang Gao. 2024. DM-CNN: Dynamic multiscale convolutional neural network with uncertainty quantification for medical image classification. *Computers in Biology and Medicine* 168: 107758. http://doi.org/10.1016/j.compbiomed.2023.107758

[7] Ana Davila, Jacinto Colan, and Yasuhisa Hasegawa. 2024. Comparison of fine-tuning strategies for transfer learning in Medical Image Classification. *Image and Vision Computing* 146: 105012. http://doi.org/10.1016/j.imavis.2024.105012

[8] Xiangzuo Huo, Gang Sun, Shengwei Tian, Yan Wang, Long Yu, Jun Long, Wendong Zhang, Aolun Li. 2024. HiFuse: Hierarchical multiscale feature Fusion Network for Medical Image Classification. *Biomedical Signal Processing and Control* 87: 105534. http://doi.org/10.1016/j.bspc.2023.105534

[9]  Ting Patrick and Kasam Anish. 2022. Applications of convolutional neural networks in chest X-ray analyses for the detection of COVID-19. *Annals of Biomedical Science and Engineering* 6, 1: 001–007. http://doi.org/10.29328/journal.abse.1001015

[10]  Prita Patil, Vaibhav Narawade. (2024). Deep Convolution Neural Network for Respiratory Diseases Detection Using Radiology Images. *International Journal of Intelligent Systems and Applications in Engineering*, 12(2), 686-704.

[11] Mahmoud Elkamouny and Milad Ghantous. 2022. Pneumonia classification for covid-19 based on machine learning. *2022 2nd International Mobile, Intelligent, and Ubiquitous Computing Conference (MIUCC)*. http://doi.org/10.1109/miucc55081.2022.9781796

[12]  Sanskruti Patel. (2021). Classification of COVID-19 from chest X-ray images using a deep convolutional neural network. *Turkish Journal of Computer and Mathematics Education (TURCOMAT), 12*(9), 2643-2651.

[13]  Prashant Patel. 2020.  Chest X-ray (Covid-19 & Pneumonia). *Kaggle*. https://www.kaggle.com/datasets/prashant268/chest-xray-covid19-pneumonia/data. Accessed on 10 May, 2024

[14]  Mohammad Afiq Hassan, Rahimi Zahari, Dina Shona Laila. (2024). COVID-19 and Pneumonia Detection System Using Deep Learning with Chest X-Ray Images. In Non-Invasive Health Systems based on Advanced Biomedical Signal and Image Processing (pp. 324-340).

[15]  Rachna Jain, Meenu Gupta, Soham Taneja & D. Jude Hemanth. (2021). Deep learning based detection and analysis of COVID-19 on chest X-ray images. *Applied Intelligence*, *51*, 1690-1700.

[16]  Sharma, A., Jha, N., & Kishor, K. (2022). Predict COVID-19 with chest X-ray. *In Proceedings of Data Analytics and Management: ICDAM 2021, Volume 1* (pp. 181-192). Singapore: Springer Nature Singapore.

[17]  Nafiz Fahad, Anik Sen, Sarzila Sahrin Jisha, Shameem Ahmad, Hazlie Mokhlis, and Md Sajid Hossain. 2023. Identification of human movement through a novel machine learning approach. *2023 Innovations in Power and Advanced Computing Technologies (i-PACT)*. http://doi.org/10.1109/i-pact58649.2023.10434296

[18]  Nafiz Fahad, Md. Kishor Morol, Anik Sen, et al. 2024. *Advancing the frontiers in credit card security: Breakthroughs in fraud detection using cutting-edge machine learning techniques*. http://doi.org/10.2139/ssrn.4801244